# Electrical control of nonlinear quantum optics in a nano-photonic waveguide

D. Hallett[1†], A. P. Foster[1*†], D. L. Hurst[1], B. Royall[1], P. Kok[1], E. Clarke[2], I. E. Itskevich[3], A. M. Fox[1], M. S. Skolnick[1] and L. R. Wilson[1]

[1]Department of Physics and Astronomy, University of Sheffield, Sheffield S3 7RH, UK

[2]Department of Electronic and Electrical Engineering, University of Sheffield, Sheffield S1 3JD, UK

[3]School of Engineering and Computer Science, University of Hull, Hull HU6 7RX, UK

\* Corresponding author: A. P. Foster: andrew.foster@sheffield.ac.uk

† These authors contributed equally to this work.

Local control of the generation and interaction of indistinguishable single photons is a key requirement for photonic quantum networks. Waveguide-based architectures, in which embedded quantum emitters act as both highly coherent single photon sources and as nonlinear elements to mediate photon-photon interactions, offer a scalable route to such networks. However, local electrical control of a quantum optical nonlinearity has yet to be demonstrated in a waveguide geometry. Here, we demonstrate local electrical tuning and switching of single photon generation and nonlinear interaction by embedding a quantum dot in a nano-photonic waveguide with enhanced light-matter interaction. A power-dependent transmission extinction as large as 40±2% and clear, voltage-controlled bunching in the photon statistics of the transmitted light demonstrate the single photon character of the nonlinearity. The deterministic nature of the nonlinearity is particularly attractive for the future realization of photonic gates for scalable nano-photonic waveguide-based quantum information processing.

Integrated quantum photonic systems, in which quantum states of light are generated, manipulated and detected on-chip, are rapidly evolving as a scalable approach for quantum information science and technologies. Efficient generation of indistinguishable single photons and control of photon-photon interactions are essential requirements for the practical implementation of integrated quantum photonics in quantum optical networks and photonic quantum computing. Integrated architectures in which embedded, waveguide-coupled quantum emitters can act as photon sources and also provide the saturable nonlinearity to mediate photon-photon interactions provide a particularly compact and scalable solution. Further key advantages of this approach include efficient coupling of the quantum emitter to a single optical mode[1] and wide bandwidth waveguide operation[2], with recent work demonstrating the significant potential of this approach using InGaAs quantum dots (QDs)[3,4] and colour centres in diamond[5,6]. The deterministic nature of the nonlinearity can be exploited in future photonic gates such as the controlled-phase[7–9] and controlled-not[10] gates, the building blocks for quantum computation using photons[11,12]. Exotic states of light, such as photon-photon bound states[13], may be derived from the nonlinear interaction. A single photon nonlinearity could also find use in classical applications such as optical routing[14] and single photon transistors[15].

Further advances in these directions require local electrical control of the waveguide-coupled quantum emitters, providing static and dynamic tuning to address challenges including emitter spectral mismatch[16] and the need for fast switching[12]. For semiconductor QD-based systems, embedding QDs in bulk Schottky or *p-i-n* structures allows for charge state selection[17,18] and ultrafast energy tuning of embedded QDs[19,20] via the DC Stark effect, as well as reducing charge noise and increasing single-photon indistinguishability[21]. Semiconductor diodes are also compatible with the generation of transform-limited single photons[22]. The challenge is to transfer similar approaches to waveguide-based integrated photonic systems. One of the key drivers here is that static electrical tuning provides a route to bringing multiple QDs into resonance, a critical step in advancing current technology beyond one and few-emitter proof-of-concept devices. On the other hand, dynamic electrical control addresses the requirement of quantum information processing for high clock speeds to limit the effects of quantum bit dephasing.

In this work, we demonstrate electrical control of resonant photon scattering from QDs in an integrated quantum photonic device, providing both spectrally tunable waveguide-coupled single photons and a switchable nonlinear response at the single photon level. Electrically tunable waveguide-coupled resonance fluorescence (RF) is measured from both neutral and charged states of the same QD. The RF signal is electrically switchable on a timescale as short as 80ns for this proof-of-concept device. In waveguide transmission measurements, we observe up to 40% extinction of a weak coherent laser tuned into resonance with the QD. Detuning-dependent bunching of the transmitted signal demonstrates the quantum nonlinear nature of the photon-emitter interaction.

**Results**

**Nano-photonic device design and characterisation**

A scanning electron microscope image of our nano-photonic device is shown in Fig. 1a. A single mode photonic crystal waveguide (PhCWG) in the centre is coupled at either end to nanobeam waveguides which are individually terminated with Bragg grating couplers (BGCs) for vertical in- and out-coupling of light. All structures are defined within a 170nm thick *p-i-n* GaAs membrane grown on a 1μm thick AlGaAs sacrificial layer (Fig. 1b). The AlGaAs layer was removed using HF wet-etching to release the GaAs membrane and maximize optical confinement within the waveguides. A layer of self-assembled InGaAs quantum dots (QDs) emitting between 880nm and 940nm was grown in the middle of the GaAs membrane. Electrical contacts fabricated on the *p* and *n* layers of the membrane enabled Stark-tuning of the QD energy states.

The PhCWG was fabricated by omitting one row of holes from a two dimensional photonic crystal. The central region of the PhCWG was designed with a photonic band edge at ~900nm (see sections S1 and S2 of the Supplementary Text for more details regarding the PhCWG design). QDs emitting near the band edge experience a Purcell enhancement due to the slow light effect[23] and couple more strongly to the single waveguide mode[24]. The BGCs have a characteristic linear optical polarisation perpendicular to the input waveguide. A 90 degree in-plane bend of one nanobeam waveguide ensured that the two BGCs were orthogonally polarized, and thus enabled efficient laser rejection during resonant transmission measurements[3].

The spectral location of the PhCWG band edge was identified using high power non-resonant photoluminescence (PL) spectroscopy. The QD ensemble within the PhCWG was excited from above and the resulting PL collected from one BGC. Figure 1c shows the resulting PL spectrum on excitation of the QD ensemble at 808nm with a power of 11μW. The band edge of the waveguide is clearly visible, with a cut-off in transmission at ~900nm. The PL spectrum is modulated by Fabry-Perot modes which are attributed to weak reflection from the BGCs.

On lowering the excitation power to 1.3μW, PL from individual QDs is observed. Figure 1d shows the PL intensity as a function of *p-i-n* junction bias (note that the large biases used in this work arise from series resistance in the contacts and have no noticeable effect on the optical properties of the device). Three bright spectral lines are observed at 893nm, 894nm and 895.5nm, labelled $X^0$, $X^+$ and $X^-$ respectively[25]. The variation in intensity of the lines with bias is indicative of charge state plateaus. Cross-correlation measurements demonstrate that all three lines originate from the same single QD (see section S3 of the Supplementary Text) and the lines are therefore attributed to different charge states of a single QD. Resonant excitation measurements with higher resolution (shown later) reveal that the line at 893nm comprises the two fine structure split states typical of $X^0$, the neutral exciton[26]. In the following, we focus on the $X^0$ and $X^-$ spectral lines.

In addition to demonstrating charge state selectivity, Fig. 1d shows that the wavelength of each spectral line is electrically tunable. A DC Stark blueshift of the QD emission of up to 0.3meV is observed as the bias is increased. For each spectral line, the Stark-tuning range is at least 5 times greater than the respective spectral width. This enables tuning of the QD into and out of resonance with a narrow linewidth probe laser, a technique which is used extensively in our measurements. For applications requiring a larger QD tuning range, we note that this can be achieved either by using QDs emitting at a longer wavelength[27] or by growing AlGaAs tunneling barriers on either side of the QDs[28].

**Electrically tunable resonance fluorescence**

Recent reports have demonstrated that resonant excitation is essential to obtain highly coherent, indistinguishable single photons from QDs[21,29,30]. With this in mind, we first demonstrate electrically tunable RF from the $X^0$ and $X^-$ spectral lines of the QD (RF was also observed from the $X^+$ spectral line, but will not be discussed here). The waveguide-coupled RF was obtained by exciting the QD from above with a continuous wave narrowband laser and collecting the scattered photons from one BGC. The RF intensity as a function of the laser wavelength and sample bias is shown in Fig. 2a, for a laser power above the objective lens of 425nW (near saturation of the QD). RF from both the $X^0$ and $X^-$ lines is observed, with a sharp transition between the two states at ~6.9V. The transition is significantly more abrupt than for non-resonant PL with above-bandgap excitation (see Fig. 1c) and is a result of the elimination of background free carriers in the vicinity of the QD when using resonant excitation. We note that the *p-i-n* diode stabilizes the QD charge environment[21] which is required to observe RF. In contrast, in undoped QD samples an additional non-resonant laser is typically needed[3,31,32].

The high resolution RF measurement reveals that the $X^0$ spectral line comprises a fine structure-split doublet. Figure 2b and 2c show RF spectra for the higher (lower) energy $X^0_1$ ($X^0_2$) fine structure states, taken from the data in Fig. 2a. The spectra were obtained by sweeping the bias whilst fixing the excitation wavelength at 893nm. Linewidths of 4.3±0.1μeV (1040±20MHz) and 3.0±0.2μeV (730±50MHz) are measured for $X^0_1$ and $X^0_2$ respectively. The fine structure splitting is ~36μeV (~8.7GHz). The contrasting intensities measured for the two states are thought to be due to the relative orientation of the orthogonal linearly polarized dipoles of the fine structure states with respect to the PhCWG, resulting in different waveguide coupling strengths (β factors) for the two states. An unfiltered resonant Hanbury-Brown and Twiss (HBT) measurement on the $X^0_1$ spectral line gave a $g^{(2)}(0)$ value of 0.16±0.04, showing that the QD scatters single photons. Figure 2d shows an RF spectrum for the $X^-$ line at a fixed excitation wavelength of 895.85nm. The linewidth of this transition is determined to be 5.1±0.1μeV (1240±30MHz).

To probe the effect of the PhCWG on the recombination dynamics of the QD, a resonant lifetime measurement was undertaken on the $X^0_1$ spectral line using pulsed excitation. The excitation and collection geometry was the same as for the continuous wave RF measurements. The lifetime was determined to be 442±3ps, corresponding to a transform limited linewidth of ~1.5μeV. For comparison, a resonant lifetime of 750±25ps was measured for the $X^0$ spectral line of a QD located in one of the nanobeam waveguides. The Purcell factor is therefore estimated to be 1.7. Additionally, a coherence time of 670±20ps was measured under low power, continuous wave resonant excitation (~40nW) for the $X_1^0$ state using Michelson interferometry. In combination with the measured lifetime this results in a long pure dephasing time of 2.8±0.1ns (see Materials and Methods). Our nano-photonic device therefore supports highly coherent QD emission, despite the proximity of the QD to numerous etched surfaces. This is likely due in part to charge stabilization provided by the *p-i-n* structure.

Figure 2e shows dynamical electrical tuning of the RF from the $X_1^0$ spectral line. A 10% to 90% switching time as short as 80ns is measured. This is significantly faster than switching by modulating the weak non-resonant laser commonly used in QD RF measurements in devices lacking a diode structure[3,32]. The switching time is limited by the large area of the diode used in this work (~0.1mm$^2$). Use of a micro-diode contacting scheme would reduce the RC time constant of the diode and allow GHz frequency modulation of the RF signal[33]. High frequency electrical control can also allow the QD transition to be locked to an external laser to enable generation of frequency-stabilized single photons[34]. This structure thus presents a promising route for sourcing of highly coherent, rapidly switchable and tunable waveguide-coupled single photons for integrated quantum optical networks.

**Waveguide transmission controlled by a single quantum dot**

The QD is a versatile nano-photonic resource. In addition to acting as a resonant single photon emitter, we demonstrate that it can also act as an electrically tunable optical nonlinear element at the single photon level. When a weak coherent laser is injected along the PhCWG, nonlinear behavior is observed because single photons are reflected by the QD whilst higher photon numbers are preferentially transmitted due to QD saturation[3]. Figure 3a shows the waveguide transmission as a function of laser wavelength and applied voltage when the laser is swept across the $X^0$ and $X^-$ states of the same QD. The transmission is normalized at each point to the transmission measured with the QD in an optically inactive state (at a bias of 5V). The power incident on the input BGC was 8.5nW, chosen to maximize the extinction of the laser by the QD. This corresponds to a power in the waveguide for which the laser field predominantly consists of single photons when measured in a time window equal to the lifetime of the QD. As the power is increased, higher photon numbers begin to dominate and the transmission extinction reduces accordingly (see section S4 of the Supplementary Text).

When the laser is resonant with the $X_1^0$ spectral line, a maximum laser extinction of 40±2% is observed. The $X_2^0$ line shows a weaker interaction with the laser (extinction up to 9±1%), consistent with the lower intensity seen for this spectral line in RF measurements. For the $X^-$ state, the maximum transmission extinction is 20±1%. The voltage dependence of the interaction of the laser and QD states is consistent with the RF measurements (see Fig. 2a), with an abrupt switch in QD charge state occurring at ~6.9V. Figure 3b and 3c show representative transmission spectra for the $X^0$ and $X^-$ spectral lines at fixed biases of 6.73V and 7.06V respectively. The dispersive Fano lineshapes are due to interference between the discrete QD spectral lines and the continuum of photonic states arising from Fabry-Perot modes formed by reflection from the BGCs and photonic crystal – nanobeam interfaces[3]. Each spectrum is fitted with a Breit-Wigner-Fano function, allowing for the probability of scattering into continuum states. Linewidths of 3.7±0.2μeV (890±50MHz) for $X_1^0$, 3.3±0.3μeV (800±70MHz) for $X_2^0$ and 4.6±0.5μeV (1100±100MHz) for $X^-$ are extracted from the fitting procedure.

The transmission contrast of 40% reported here is significantly higher than that measured in single-pass waveguide structures without electrical control[3,6], and comparable with a cavity-based approach[5] but with the highly desirable property of broadband waveguide coupling. Deterministic coupling of single photons with the QD is required for many future device applications. To identify the factors which currently limit the transmission contrast we model the device using the transfer matrix method (see Supplementary Text section S5 for full details). Figure 3d shows the results of the model alongside the experimental transmission as a function of the $X_1^0$-laser detuning. The degree of QD blinking was used as the only fitting parameter in the model, with a value of ~9% describing the data well. This is in agreement with the absence of blinking noted on the second timescale of the transmission measurements. From the model we conclude that approaching the transform limit for the QD linewidth and increasing the Purcell factor to 5 would result in over 90% transmission extinction. These values are achievable with current technological capabilities. The linewidth of the $X_1^0$ spectral line studied here is already only a factor of 2.5 greater than the transform limit, but could be improved with greater control over charge and spin noise in the vicinity of the QD[22]. A larger Purcell enhancement should be achievable in the waveguide geometry, with values up to 30 predicted theoretically for slow light PhCWGs[24].

The extinction of the laser by the QD can be switched rapidly using electrical tuning in the same manner as the RF signal. Figure 3e shows the switching response of the device when the $X_1^0$ line is tuned out of resonance with the laser. The transmission state is switched in as little as 60ns, in agreement with the RF switching time. This measurement demonstrates that the $X_1^0$ spectral line can therefore be used to switch the transmission state in the current device at a frequency of ~10MHz. As noted above in the case of RF, optimization of the diode structure would enable GHz switching speeds[33].

**Transmission photon statistics and optical nonlinearity**

The quantum nature of the nonlinear photon-QD interaction was confirmed using photon statistics measurements on the transmitted laser field, with the laser resonant with the $X_1^0$ spectral line. The laser power incident on the BGC was 8.5nW, and the transmission dip on resonance was 20% (reduced relative to Fig. 2 as the alignment of input and output polarizers was optimized for maximum photon detection count rate rather than transmission contrast). Figure 4a shows the resulting histogram of correlated events binned by time delay. Clear bunching of 1.14±0.01 is observed on the timescale of the $X_1^0$ lifetime. This indicates that there is an increased probability of transmission when two or more photons interact with the QD within the $X_1^0$ lifetime, and a reduced probability of transmission of single photons. The degree of bunching is greater than that measured previously for a QD in a PhCWG without electrical charge stabilization[3].

Figure 4b shows the degree of bunching as a function of the $X_1^0$-laser detuning. The detuning is controlled by fixing the laser wavelength and electrically tuning the $X_1^0$ transition. At large detuning values $g^{(2)}(0)$ approaches unity, as expected for a coherent laser source described by Poissonian statistics. The value of $g^{(2)}(0)$ increases from unity as the detuning is reduced and reaches a maximum of 1.14 for zero detuning. The width of the detuning curve is close to the 3.7μeV spectral width of the $X_1^0$ transition (shown as a dashed Lorentzian curve), clearly demonstrating that the degree of bunching is dependent on the strength of the interaction between the laser and the QD.

**Discussion**

In conclusion, we have demonstrated electrical control of resonant scattering from a single QD embedded in a nano-photonic waveguide. The QD was shown to act both as a source of electrically tunable, waveguide-coupled single photons and as a nonlinear element to control photon-photon interactions in the waveguide. High purity single photon emission was demonstrated under resonant excitation with $g^{(2)}(0)=0.16$. The transmission of a weak coherent laser through the waveguide was modulated by up to 40% by electrically controlling the QD-laser detuning. Detuning dependent bunching of the transmitted photons revealed the quantum nature of the nonlinearity.

Achieving complete transmission extinction is highly desirable, resulting in a pi phase shift being applied to every scattered single photon (which follows from Ref. 35). The narrow linewidths reported here for a QD embedded in a photonic crystal waveguide represent very encouraging progress towards the transform limited linewidths which complete transmission extinction requires. Electrical control of the quantum optical nonlinearity opens up several promising avenues for future research. In particular, many quantum optical devices require two or more indistinguishable emitters. This work demonstrates that the electrical approach to QD energy tuning (required to overcome spectral mismatch between QDs) is fully compatible with the realisation of a single photon nonlinearity in a waveguide-coupled geometry. Electrical tuning is a local approach[36] and will allow for individual electrical control of separate QDs coupled to the same waveguide mode. One can also envisage a realisation of the quantum optical controlled-phase gate[7], containing two QDs tuned into resonance via the DC Stark effect.

## Materials and Methods

### Sample fabrication

The sample was grown using molecular beam epitaxy with the epitaxial layer structure given in Fig. 1b. A 170nm thick GaAs *p-i-n* membrane containing self-assembled InGaAs QDs in the intrinsic region was grown on top of a 1µm *n*-$Al_{0.6}Ga_{0.4}As$ sacrificial layer. The 30nm thick *p* and *n* layers of the membrane had dopant densities of $2\times10^{19}cm^{-2}$ and $2\times10^{18}cm^{-2}$ respectively. Devices were patterned using electron beam lithography and transferred into the membrane by inductively coupled plasma etching. Ti:Au contacts were formed to the *p* and *n* layers of the membrane to complete the diode structure. Finally, a wet etch with 40% HF released the devices.

### Experimental setup

Optical measurements were undertaken with the sample mounted in a home-made gas exchange cryostat operating at 4.2K. The cryostat had optical access from above the sample, which was positioned using ultrastable x, y, and z piezo-stages. Independent spatial selectivity of excitation and collection locations was enabled using fiber-coupled optics external to the cryostat.

### Resonant excitation techniques

Continuous wave (CW) resonant optical measurements were performed using a narrow band, widely tunable CW Ti:Sapphire laser (SolsTiS-1000-SRX-XF-TS3, M Squared Lasers). Rejection of the scattered laser background in resonant measurements was achieved using a combination of spatial and polarisation filtering. A differential technique was also employed to remove residual background laser scatter. This entailed electrically switching the QD into and out of resonance with the laser at a frequency of 1kHz-10kHz, and evaluating the difference signal ('resonant' minus 'off resonant'). A signal to noise ratio (SNR) of up to 200:1 was obtained in the case of CW RF, upon exciting a QD in the PhCWG from above and collecting the resulting waveguide-coupled scattered photons from one of the BGCs.

For Hanbury-Brown and Twiss measurements, the transmitted light was split using a 50:50 fiber beam splitter and detected by two avalanche photodiodes (APDs). Correlations between detection events were analysed using a time-correlated single photon counting card (Becker-Hickl SPC-130).

Resonant lifetime measurements were undertaken using a picosecond pulsed Ti:Sapphire laser (Spectra-Physics Tsunami, Newport). A monochromator was used to create a narrow bandwidth excitation pulse with a width of ~10pm. The QD was excited from above and scattered photons were collected from one BGC and detected on a superconducting nanowire single photon detector with a time resolution of 50ps (Single Quantum Eos). The SPC-130 single photon counting card was used to correlate the arrival times of scattered photons with a reference pulse from a photodiode. To account for background scatter, the QD was electrically modulated on and off resonance with the laser. Events detected in the off resonant case were removed from the data obtained in the resonant case, thus removing background scatter from the measurement.

To determine the QD coherence time, photons scattered resonantly from the QD under CW excitation were passed through a free space Michelson interferometer, one arm of which contained a variable delay line. The interference signal as a function of time delay between the two arms was read out using an APD. The pure dephasing time was calculated from the well-known expression $1/T_2 = 1/2T_1 + 1/T_2^*$ where $T_1$ and $T_2$ are the QD lifetime and coherence time respectively, and $T_2^*$ is the pure dephasing time.

one-dimensional nanophotonic waveguides coupled to a qubit. *Phys. Rev. A - At. Mol. Opt. Phys.* **82,** 063821 (2010).

36. Bentham, C. *et al.* Single-photon electroluminescence for on-chip quantum networks. *Appl. Phys. Lett.* **109,** 161101 (2016).
**Acknowledgments:**

**Funding:** This work was funded by EPSRC grant number EP/N031776/1. **Author contributions:** The sample was made by B.R. and E.C. The experiments were performed by D.H. and A.P.F. L.R.W, M.S.S, I.E.I. and A.M.F conceived the experiment and guided the work. The theoretical model was created by D.L.H. and P.K. The paper was written by A.P.F and D.H. with contributions from all authors. **Competing interests:** The authors declare that they have no competing interests.
17

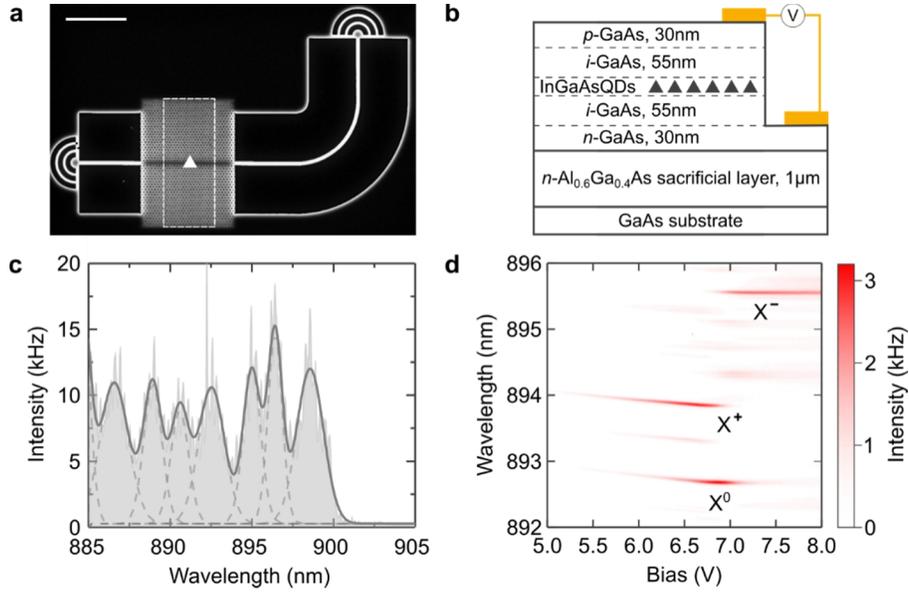

**Fig. 1. Device structure and optical characterization. (a)** Scanning electron microscope image of the device. The white dashed box encloses the slow light section of the PhCWG. The triangle shows the approximate location of the studied QD. Scale bar 5μm. **(b)** Diode structure schematic. Electrical contacts are made to the *p*- and *n*-GaAs layers. **(c)** High power photoluminescence spectrum at a bias of 8V showing emission from an ensemble of QDs in the photonic crystal waveguide (light grey shaded region). The waveguide band edge is seen at ~900nm. Fabry-Perot oscillations are observed due to reflection from the BGCs. The peaks are fit with multiple Gaussians (dashed lines) with their sum given by the dark grey line. **(d)** Photoluminescence intensity versus wavelength and bias for non-resonant excitation in the centre of the photonic crystal waveguide and collection from one BGC.

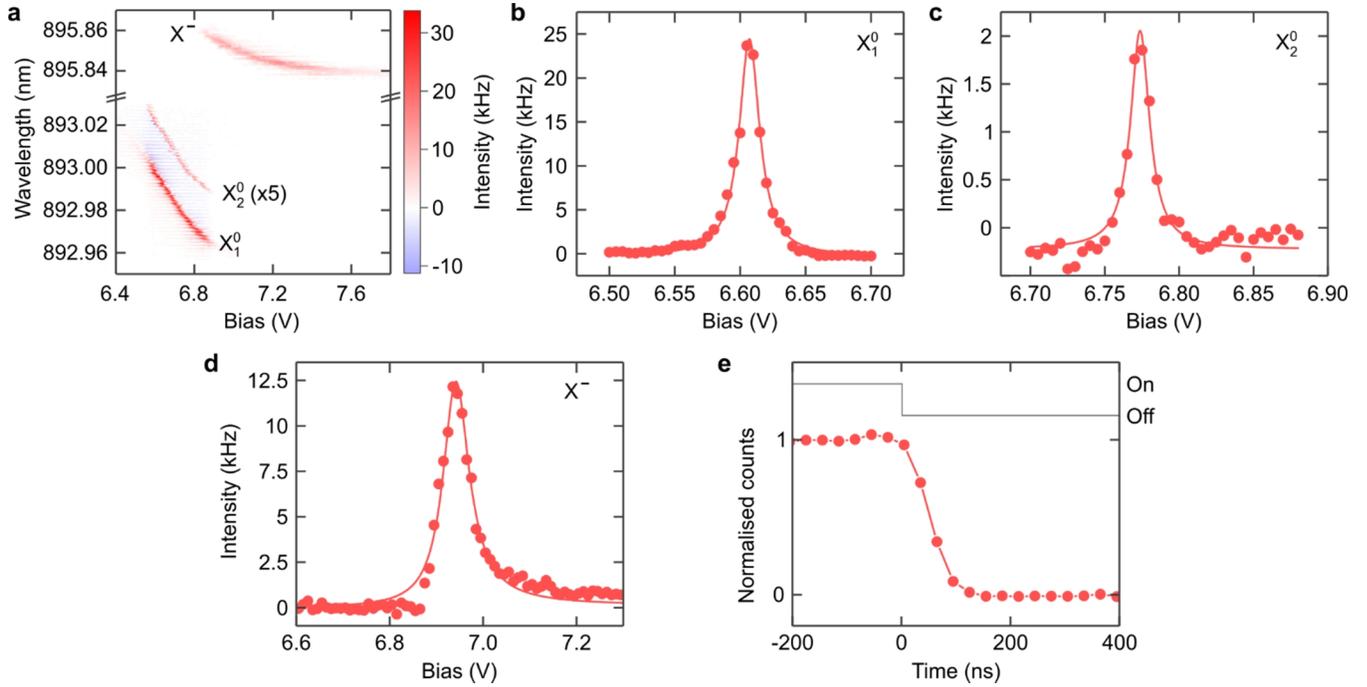

**Fig. 2. Resonance fluorescence (RF) from a QD in the photonic crystal waveguide. (a)** RF intensity as a function of wavelength and bias for the neutral fine structure states ($X_1^0$ and $X_2^0$) and charged state ($X^-$) of a single QD. The $X_2^0$ state intensity has been scaled by a factor of 5 for clarity. **(b, c)** Swept-bias RF spectra (circles) for the **(b)** higher energy $X_1^0$ state and **(c)** lower energy $X_2^0$ state of the neutral exciton at a fixed excitation wavelength of λ=893nm. **(d)** Swept-bias RF spectrum (circles) for the $X^-$ state of the same QD at a fixed wavelength of 895.85nm. Solid lines are Lorentzian fits to the data. **(e)** Electrical switching of the RF from the $X_1^0$ state for λ=893nm (circles; the lower red line is a guide for the eye). The upper grey line indicates the $X_1^0$-laser detuning during the measurement ('on' – resonant, 'off' – non-resonant).

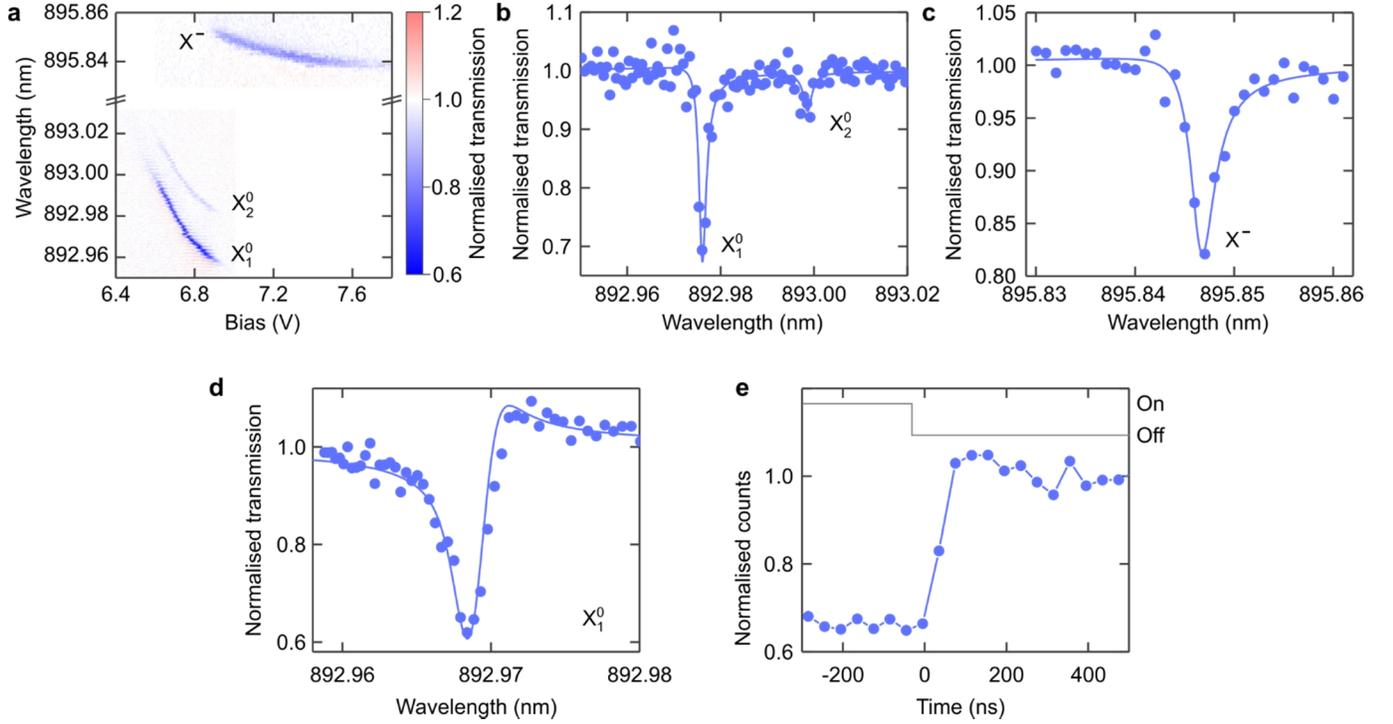

**Fig. 3. Resonant transmission through the photonic crystal waveguide. (a)** Normalised transmitted intensity as a function of wavelength and bias for the neutral fine structure states ($X_1^0$ and $X_2^0$) and charged state ($X^-$) of a single QD. **(b, c)** Normalised transmission spectra (circles) for the **(b)** $X_1^0$ and $X_2^0$ states of the neutral exciton at a bias of 6.73V and **(c)** the $X^-$ state of the same QD at 7.06V. Solid lines are Breit-Wigner-Fano fits to the data. **(d)** Transmission extinction measured on resonance with the $X_1^0$ spectral line (circles). The experimental data was acquired by sweeping the bias at fixed wavelength (λ=892.97nm). The bias was then converted to wavelength using the voltage-wavelength dependence determined from **(a)**. This removes any wavelength-dependent variation in the scattered laser suppression. It is also the reason for the mirrored shape of the Fano resonance compared with the data in **(b)**. The solid line shows the transmission predicted by a transfer matrix model (see main text and Supplementary Text.). **(e)** Electrical switching of transmission by the $X_1^0$ spectral line (circles; the lower blue line is a guide for the eye). The upper grey line indicates the $X_1^0$-laser detuning during the measurement ('on' – resonant, 'off' – non-resonant).

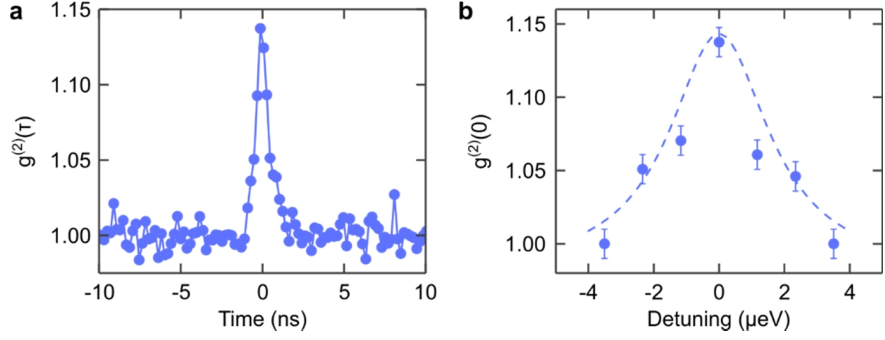

**Fig. 4. Photon statistics for transmitted photons resonant with the $X_1^0$ spectral line.** **(a)** Second order autocorrelation function for transmitted photons at zero $X_1^0$-laser detuning. Bunching is observed on the timescale of the lifetime of the $X_1^0$ state. **(b)** Maximum bunching versus $X_1^0$-laser detuning (circles). Error bars equal one standard deviation in the noise level. The dashed line is a Lorentzian with the $X_1^0$ transmission linewidth of 3.7μeV as a guide to the eye.